\documentclass[twocolumn,rmp,showpacs]{revtex4}

\usepackage{dcolumn,graphicx,amsmath,amssymb}
\usepackage{txfonts}

\newcommand{\ccp}{c_{\mathrm{cp}}}
\newcommand{\dr}{\Delta r}

\begin{document}

\title{Core-periphery organization of complex networks}
\author{Petter Holme}
\affiliation{Department of Physics, University of Michigan, Ann Arbor,
MI 48109, U.S.A.}

\begin{abstract}
  Networks may, or may not, be wired to have a core that is both
  itself densely connected and central in terms of graph distance. In
  this study we propose a coefficient to measure if the network has
  such a clear-cut core-periphery dichotomy. We measure this
  coefficient for a number of real-world and model networks and find
  that different classes of networks have their characteristic
  values. For example do geographical networks have a strong
  core-periphery structure, while the core-periphery structure of
  social networks (despite their positive degree-degree correlations)
  is rather weak. We proceed to study radial statistics of the core,
  i.e.\ properties of the $n$-neighborhoods of the core vertices for
  increasing $n$. We find that almost all networks have unexpectedly
  many edges within $n$-neighborhoods at a certain distance from the
  core suggesting an effective radius for non-trivial network processes.
\end{abstract}

\pacs{89.75.Fb, 89.75.Hc}

\maketitle

\section{Introduction}

All systems consisting of pairwise-interacting entities can be
modeled as networks. This makes the study of complex networks one of
the most general and interdisciplinary areas of statistical
physics~\cite{mejn:rev,ba:rev,doromen:book}. One of the most important
gains of the recent wave of statistical network studies is the
quantification of large-scale network
topology~\cite{luc:rev,mejn:rev}. Now, with the use one just a few
words and numbers, one can state the essential characteristics of a
huge network---characteristics that also say something about how
dynamical systems confined to the network will behave.
A possible large-scale design principle is that one part of the network
constitutes a densely connected core that also is central in terms of
network distance, and the rest of the network forms a periphery. In,
for example, a network of airline connections you would most certainly
pass such a core-airport on any many-flight itinerary. It is
known that a broad degree distribution can create a core having these
properties~\cite{chung_lu:pnas}. In this paper we
address the question if there is a tendency for such a structure in the
actual wiring of the network. I.e., if one assumes degree to be,
to a large extent, an intrinsic property of the vertices, then is
the network organized with a distinct core-periphery structure or not?
To give a quantitative answer to this question our first step is to
find a core with the above mentioned properties---being
highly-interconnected and having a high closeness
centrality~\cite{sab:clo} (the inverse average distance between a
vertex in the core and an arbitrary vertex). Once such a subgraph is
identified we calculate its closeness centrality relative to the graph
as a whole, and subtract the corresponding quantity for the ensemble of
random graphs with the same set of degrees as the original network
(cf.\ Ref.~\cite{maslov:pro}). If the resulting coefficient is
positive the network shows a pronounced core-periphery structure.
Once the core and periphery are distinguished one may proceed to
investigate their structure. We look at the statistical properties of
the $n$-neighborhoods (the set of vertices on distance $n$) of the
core vertices. By such radial statistics we can draw conclusions for
the respective function of the core and periphery.
This paper starts by defining the core-periphery coefficient and
measure it for real-world networks of numerous types, then proceeds
by discussing and measuring radial statistics.

\begin{table*}
\caption{\label{tab:ccp} The network sizes $N$ and $M$, the
  core-periphery coefficient $\ccp$ and the relative assortative
  mixing coefficient $\dr$ for a number of networks. In the
  interstate network the vertices are American interstate highway
  junctions and two junctions are connected if there is a road with no
  junction in between. The pipeline network is a similar network of
  junctions and gas pipes. In the street networks the vertices are
  Swedish city street segments connected if they share a junction. In
  the airport data (obtained from IATA www.iata.org) the
  vertices are airports and the edges represent airport pairs with a
  non-stop flight connection. The Internet figures are averages of 15
  AS-level graphs constructed from traceroute searches. The arXiv,
  board of directors and Ajou students are constructed one-mode
  projections from affiliation networks (where links goes from persons
  to e-prints, corporate boards and university classes
  respectively). The student network is averaged over graphs for 16
  semesters. One edge represent two students taking at least three
  classes together that semester. The high school, prisoner and social
  scientist networks are gathered from questionnaires---an edge means
  that two
  persons have listed each other as acquaintances. The high school
  data are averaged over 84 individual schools. In the electronic
  communication networks one edge represent that at least one of the
  vertices has contacted the other over some electronic medium. In the
  nd.edu data the vertices are HTML documents and the edges are
  hyperlink. The citation graph is constructed from preprints in the
  field of high-energy physics~\cite{kdd:hep} (see:
  http://www.cs.cornell.edu/projects/kddcup/datasets.html).
  In the software dependency graph the vertices are software packages
  and an edge means that one package needs the other for its proper
  function. The food webs are networks of water-living species and an
  edge means that one species prey on the
  other. For the protein networks an edge means that two proteins bind
  to each other physically. The metabolic and ``whole cellular''
  networks consist of chemical substances and edges indicating that
  one molecule occur in the same reaction as the other (the values for
  these networks are averages over 43 organisms from different domains
  of life).
}
\begin{ruledtabular}
  \begin{tabular}{lrc|cccc}
   \multicolumn{2}{c}{Network} & Ref. & $N$ & $M$ & $\ccp$ & $\dr$\\\hline
    Geographical networks & Interstate highways & & 935 & 1315 &
    0.231(1) & 0.0851(5)\\
    & Pipelines & \cite{gast:effi} & 2999 & 3079 & 0.180(2) & 0.073(2)\\
    & Streets, Stockholm & \cite{rosv:city} & 3325 & 5100 & 0.255(1) &
    0.080(1)\\
    & Streets, G\"{o}teborg  & \cite{rosv:city} & 1258 & 1516 &
    0.040(3) & 0.019(3)\\
    & Airport & & 449 & 2795 & 0.0523(3) & 0.0910(3)\\
    & Internet & \cite{vesp:inet} & 1968(66) & 4051(121) & 0.045(2) &
    0.009(3)\\\hline
    One-mode projections of & arXiv & \cite{mejn:scicolpre1} & 48561 &
    287570 & --0.08(3) & 0.361(3)\\
    affiliation networks & Board of directors & \cite{davis} & 6193 &
    43074 & --0.037(2) & 0.280(2)\\ 
    & Ajou University students & \cite{our:ajou1,our:ajou2}& 7285(128)
    & 75898(6566)& --0.08(1) & 0.66(4)\\\hline
    Acquaintance networks & High School friendship & \cite{addh} &
    571(43) & 1078(85) & 0.006(7) & 0.19(1)\\
    & Prisoners & \cite{mcrae:prison} & 58 & 83 & --0.043(2) &
    0.264(2)\\
    & Social scientists & \cite{freeman:eiec} & 34 & 265(35) &
    -0.002(4) & 0.10(1)\\\hline
    Electronic communication & e-mail, Ebel \textit{et al.} &
    \cite{bornholdt:email} & 39592 & 57703 & --0.229(4) & --0.001(4)\\
    & e-mail, Eckmann \textit{et al.} & \cite{eckmann:dialog} & 3186 &
    31856 & --0.091(2) & --0.034(2)\\
    & Internet community, nioki.com & \cite{nioki} & 49801 & 239265 &
    --0.014(2) & 0.007(2)\\
    & Internet community, pussokram.com & \cite{pok} & 28295 & 115335
    & --0.183(5) & --0.005(5)\\\hline
    Reference networks & WWW, nd.edu & \cite{ba:dia} & 325729 &
    1090108 & --0.027(3)& --0.003(3)\\
    & HEP citations & & 27400 & 352021 & --0.10(1) & 0.03(1) \\\hline
    Software dependencies & GNU / Linux & \cite{mejn:mix} & 504 & 793
    & --0.155(1) & --0.087(1)\\\hline
    Food webs & Little Rock Lake & \cite{martinez:rock} & 92 & 960 &
    0.005(6) & --0.0141(6)\\
    & Ythan Estuary & \cite{ythan1} &  134 & 593 & --0.020(1) &
    --0.0153(9)\\\hline
    Neural network & \textit{C.\ elegans} & \cite{cenn:brenner} & 280
    & 1973 & 0.040(6) & 0.0222(7)\\\hline
    Biochemical networks & \textit{Drosophila} protein &
    \cite{droso:pin} & 2915 & 4121 &
    --0.035(2) & 0.003(1)\\
    & \textit{S.\ cervisiae} protein & \cite{pagel:mips} & 3898 & 7283
    & --0.249(1) & --0.069(1)\\
    & \textit{S.\ cervisiae} genetic & \cite{pagel:mips} & 1503 & 5043
    & --0.0646(7) & --0.101(1)\\
    & Metabolic networks & \cite{jeong:meta} & 427(27) & 1257(88) &
    --0.002(6) & 0.006(1)\\
    & Whole cellular networks & \cite{jeong:meta} & 623(32) &
    1752(103) & --0.004(6) & --0.001(2)
\end{tabular}
\end{ruledtabular}
\end{table*}

\section{Measuring the core-periphery structure}

In this paper we assume the network to be represented as a graph
$G=(V,E)$ with a set $V$ of $N$ vertices and a set $E$ of $M$
undirected and unweighted edges. (It is straightforward to generalize
our analysis to weighted networks.) Since our analysis requires the
network to be connected we will henceforth identify $G$ with the
largest connected component of the network (in all mentioned cases
this component will constitute almost the entire network). We also
remove self-edges and multiple edges.

\subsection{Rationale and definition of the core-periphery
  coefficient}

The notion of network centrality is a very broad and many measures
have been proposed to capture different aspects of the
concept~\cite{harary}. One of the simplest quantities is the closeness
centrality~\cite{sab:clo}
\begin{equation}\label{eq:cc_vertex}
  C_C(i) = \left(\langle d(i,j) \rangle_{j\in
  V\setminus\{i\}}\right)^{-1}
\end{equation}
of a vertex $i$, where $d(i,j)$ is the distance between $i$ and $j$ (the
smallest number of edges on a path from $i$ to $j$). The closeness of
a vertex is thus the reciprocal average shortest distance to the
other vertices of $V$. This definition is straightforwardly extended
to a subset $U$ of vertices
\begin{equation}\label{eq:cc_vertex}
  C_C(U) = \left(\left\langle \langle d(i,j) \rangle_{j\in
  V\setminus\{i\}}\right\rangle_{i\in U}\right)^{-1} .
\end{equation}
So we require a core to be a subgraph $U$ with high $C_C(U)$, but also
to be a well-defined cluster---i.e.\ to have comparatively many edges
within. Now, if there are many facets of the centrality concept, there
are even more algorithms to identify graph clusters, each being a
\textit{de facto} cluster-definition~\cite{mejn:rev}. For
simplicity we choose the most rudimentary cluster-definition---the set of
$k$-cores. A $k$-core is a maximal subgraph with the minimum degree
$k$ (maximal in the sense that if one adds any vertex to a $k$-core it
will no longer have a minimal degree $k$). To calculate a sequence of
$k$-cores is computationally cheaper (linear
in $M$~\cite{rama:core}) than more elaborate clustering
algorithms.\footnote{One iteratively removes the vertex of currently
  lowest degree $k_\mathrm{min}$, if $k_\mathrm{min}$ is not lower
  than its largest value during the iterations then the remaining
  network is a $k$-core.}
So we let our core $V_\mathrm{core}(G)$ be the $k$-core with maximal
closeness and define the core-periphery coefficient $\ccp$ as
\begin{equation}\label{eq:cpc}
  \ccp(G) = \frac{C_C[V_\mathrm{core}(G)]}{C_C[V(G)]} -
  \left\langle\frac{C_C[V_\mathrm{core}(G')]}
  {C_C[V(G')]}\right\rangle_{G'\in\mathcal{G}(G)},
\end{equation}
where $\mathcal{G}(G)$ is the ensemble of graphs with the same set of
degrees as $G$. The sequence of $k$-cores is not necessarily
unique. We maximize $C_C(U)$ over $m_\mathrm{seq}$ different
sequences. In practice different runs almost always yield the same
core, and the value of $m_\mathrm{seq}$ seems to matter
little. The $m_\mathrm{null}$ elements of $\mathcal{G}(G)$ can be
obtained by randomization of $G$ in time and space of the order of
$M$~\cite{roberts:mcmc}. In this paper we use $m_\mathrm{null}=1000$
and $m_\mathrm{seq}=10$ for networks with $N<5000$, and
$m_\mathrm{null}=50$ and $m_\mathrm{seq}=3$ for $N\geq 5000$.

The correlation of degrees at either side of an edge is an informative
structure to
study~\cite{maslov:pro,mejn:assmix,pas:inet}. To some extent one can
see degree-degree correlations as a local version of the
core-periphery structure---when there are positive degree-degree
correlations, at least some subgraphs of the
network will have a well-defined core and periphery. Such clusters
need not be centrally positioned in the graph as a whole, so while the
degree-degree correlations says something about if the graph can be
separated into densely and sparsely connected regions, the
core-periphery structure gives information of the relative position of
such regions. A common way to quantify the average degree-degree
correlations is to measure the \textit{assortative mixing
coefficient}~\cite{mejn:assmix}
\begin{equation}\label{eq:r}
  r=\frac{4\langle k_1\, k_2\rangle - \langle k_1 + k_2\rangle^2}
  {2\langle k_1^2+k_2^2\rangle - \langle k_1+ k_2\rangle^2} ,
\end{equation}
where $k_i$ is the degree of the $i$'th argument of a edge as it
appear in a list of $E$. Now, our null-model is a random graph
conditioned to have the same degree sequence as the original
graph. In other words, just as for the core-periphery structure, we
consider the deviation from our null model and measure the relative
assortative mixing coefficient
\begin{equation}\label{eq:relr}
  \dr(G)= r(G)-\left\langle r(G')\right\rangle_{G'\in\mathcal{G}(G)}.
\end{equation}

\subsection{Numerical results for real-world networks}

In Table~\ref{tab:ccp} $\ccp$ and $\dr$ are displayed for a number
of real-world networks. We find that the core-periphery structure and
relative degree-degree correlations follow the different classes of
networks rather closely. Furthermore the core-periphery
structure and degree-degree correlations seem to be quite independent
network structures in practice. For example,
geographically embedded networks have a clear core-periphery structure
and weakly positive degree-degree correlations; social networks
derived from affiliations have slightly negative $\ccp$-values but
very high $\dr$-values; networks of online communication have markedly
negative $\ccp$ and rather neutral degree-degree correlations.
Most geographically embedded networks have the function of
transporting, or transmitting, something between the vertices. 
 Networks with a well-defined core (which
most paths pass through) and a periphery (covering most of the area)
are known to have good performance with respect to communication
times~\cite{gast:effi}. Also networks of airline traffic~\cite{gui:air} and the
hardwired Internet~\cite{zhou:rich,vesp:inet} are
known to have well-defined cores due to traffic-flow optimization.
The class of one-mode projection networks (social networks constructed
by linking people that participate in something---movies, scientific
research, etc.---together) show slightly negative $\ccp$-values. This
can, at least for the data sets of scientific
coauthors~\cite{mejn:scicolpre1} and fellow students of a Korean
university~\cite{our:ajou1,our:ajou2}, be explained by that there is a
grouping of the people on the basis of specialization (and, in student
networks, also in grade) and thus no well-defined core. We note that
this group of networks have very high $\dr$ values.
The interview based acquaintance networks show rather neutral
$\ccp$-values and positive $\dr$ suggesting that there is a degree of
independence between. This is quite similar to the one-mode
projections, which probably is not a coincidence---there is a strong
correlation between acquaintance ties and the organizations people are
affiliated with.
The vertices in electronic communication networks are also people but
the network structures of these are quite different; the degree-degree
correlation is typically slightly negative, as is the
core-periphery coefficient.
Information networks where the edges refer to supporting information
sources (our examples are a subgraph of the WWW and a graph of citations
between papers in high-energy physics) can be expected to be grouped
into topics, thus the negative $\ccp$.
The same explanation applies to the negative $\ccp$ of the software
dependency graph.
Food webs are other stratified networks where a lack of a well-defined
core seems natural.
The core of the neural network of \textit{C.\ elegans} is a clique
(fully connected subgraph) of eight neurons, which accounts for
positive $\ccp$ and $\dr$ values.
The biochemical networks all show negative $\ccp$ values and negative,
or neutral, relative assortative mixing coefficients.

\subsection{Numerical results for network models}

\begin{figure}
  \resizebox*{0.85 \linewidth}{!}{\includegraphics{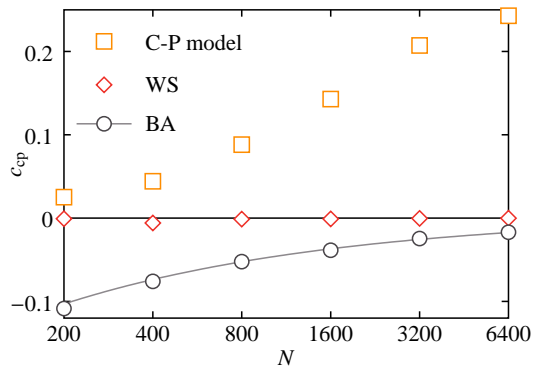}}
  \caption{ Core-periphery structure of model networks. The
    Barab\'{a}si-Albert and Watts-Strogatz networks have $M=2N$. The
    core-periphery model has the parameter $f_\mathrm{core}=0.96$
    (i.e.\ the intended core consists of 4\% of the vertices) and
    $\gamma = 3$. All values are averaged over $10^4$-$10^5$ network
    realizations. The BA-model line is a fit to an power-law form
    $\alpha_0+\alpha_1N^{-\alpha_2}$ (this fit gives
    $\ccp(\infty)=\alpha_0=0.004(9)$).
  }
  \label{fig:mod}
\end{figure}

In addition to the real-world networks of Table~\ref{tab:ccp} we also
measure the core-periphery coefficient for a few network models. 
For simple random graphs~\cite{er:on} where $N$ vertices are randomly
connected by $M$ edges, defining an ensemble $\mathbb{G}(N,M)$ of
graphs, $\mathcal{G}(G)$ is precisely the elements of
$\mathbb{G}(N,M)$ with the same degree sequence as $G$. This means
that, on average, $\ccp$ will be zero for random graphs.
A popular network model is the Barab\'{a}si-Albert (BA)
model~\cite{ba:model} where the graphs are grown by iteratively adding
new vertices with edges to old vertices with a probability
proportional to the degree of the old vertices. In Fig.~\ref{fig:mod}
we see that $\ccp$ tends to zero (or a value very close to zero) for
BA model networks. The BA model has an assortative mixing coefficient
$r$ that tends to zero
as $N$ grows~\cite{mejn:assmix}. From this one sees that the
high-degree vertices are not more interconnected than can be expected
from their degrees, and thus that there is no preference in the actual
wiring of the network for a well-defined core in the sense of the
$\ccp$-coefficient. 
We also investigate the Watts-Strogatz' small-world network
model~\cite{wattsstrogatz} were one end of the edges of a
circulant~\cite{harary} is rewired with a certain probability ($0.01$
in our case). Just as for the BA model $\ccp$ converges to zero (see
Fig.~\ref{fig:mod}). This is not so surprising, in the WS model's
starting point, the circulant, every vertex is in the same
position. The rewiring procedure does not aggregate vertices to a
well-defined core either.
Finally we construct a network model with a positive core-periphery
coefficient. We start by drawing $N$ power-law distributed random
integers in the interval $[2,\infty)$, i.e.\ the probability for a
number $m$ to be drawn is proportional to $m^{-\gamma}$, and sort
these numbers in increasing order: $m_1,\cdots,m_N$. These numbers are
the desired degrees of the vertices and can be thought of as stubs, or
half-edges, that need to be connected. Now we will attempt to make a
well-defined core of the vertices $i_1^\mathrm{core},\cdots,N$, where
$i_1^\mathrm{core}$ is the
integer closest to $Nf_\mathrm{core}$ (so $f_\mathrm{core}$ is a
parameter setting the relative size of the core). Then we go through
the vertices $i_1^\mathrm{core},\cdots,N$ in increasing order and for each vertex
$i$ try to attach the stubs the vertices $j=i+1,\cdots,N$ (once again in
increasing order) as long as the degree of $j$ is less than $m_j$.
The remaining stubs are paired together randomly and made into edges
if they do not form loops or multiple edges. The superfluous stubs are
then deleted. For this model $\ccp$ indeed shows positive and growing
values, see Fig.~\ref{fig:mod}.

\section{Radial organization of networks}

\begin{figure*}
  \resizebox*{0.95 \linewidth}{!}{\includegraphics{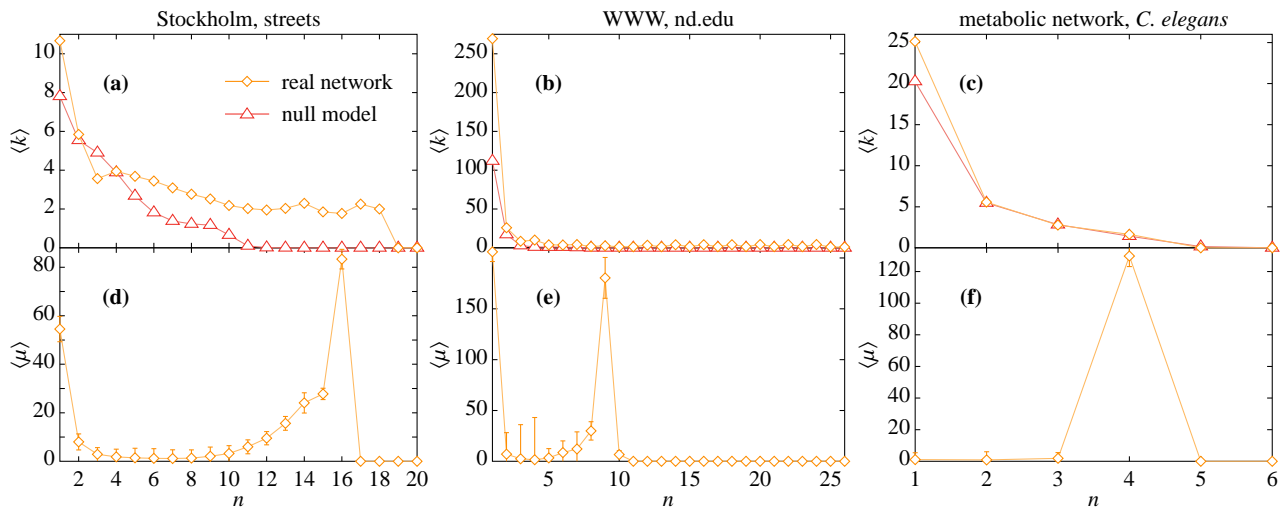}}
  \caption{ Radial statistics for three real-world networks. (a)-(c)
    show the average degree $\langle k\rangle$ of the
    $n$-neighborhoods of the core vertices as a function of $n$ for
    three real world networks: a network of streets in Stockholm,
    Sweden~\cite{rosv:city}, a network of hyperlinked
    web-pages~\cite{ba:dia} and the metabolic network of \textit{C.\
    elegans}~\cite{jeong:meta}. Curves for our null-model---random
    networks with the same degree sequence as the original
    network---are included. In (a)-(c) we plot the number of edges
    with in the $n$-neighborhood relative to the expected number of
    edges given the degree sequence of the $n$-neighborhood and the
    graph as a whole. Lines are guides for the eyes.
  }
  \label{fig:rad}
\end{figure*}

A well-defined core is a useful starting point
for a radial examination of the network. By plotting quantities
averaged over the $n$-neighborhoods (the set of vertices at a distance
$n$ of a vertex) of the core vertices as functions of $n$ one
can get an idea of the respective purposes of the core and
periphery. This kind of statistics is naturally more sensible the
stronger the core-periphery structure is. The $\ccp$ construction
identifies the most central well-connected core but it does not
say whether or not the core make sense---even for slightly negative
$\ccp$-values this type of radial statistics may be informative.
While authors have focused on the size of the $n$-neighborhoods of
random vertices~\cite{radek,cohen:tomo}---a useful approach to
monitor finite-size effects that affects spreading processes such as
disease epidemics---we will focus on quantities that we find more
informative regarding the relative functions of the core and
periphery.

To get a rough view of the radial network organization we plot the
average degree of the vertices in the $n$-neighborhood of
core-vertices as a function of $n$ in Fig.~\ref{fig:rad}. We include
the corresponding results for our null model (random networks
constrained to the same degree sequence as the original). The core
vertices themselves almost always get higher average degree for the
null-model than the real-world networks ($5$-$10\%$ higher for the
networks of Fig.~\ref{fig:rad}). For the first neighborhood the
situation is reversed---the real-world networks have higher $\langle
k\rangle$ than the null-model. Then the degrees are decreasing
monotonically; typically faster for the null-model networks. For the
street network in Fig.~\ref{fig:rad}(a) $\langle k\rangle$ decreases
rather slowly for intermediate $n$; the periphery is thus rather
homogeneous. The short average distances of the core, consisting of
the streets of the city center, can be attributed to its central
geographic position. 

One can imagine different functions of the peripheral
vertices---either they are just conveying information, traffic, etc.\
to and from the core; or they are, just as the core-vertices, involved
in the general network processes, only less intensely. To understand
this we measure the average value of the quantity
\begin{equation}\label{eq:phi}
  \mu(i,n) = M(K_n(i)) / \mathbb{E} M(K_n(i))
\end{equation}
over the core vertices; $M(K_n(i))$ is the number of edges within
$i$'s $n$-neighborhood $K_n(i)$ and $\mathbb{E} M(K_n(i))$ is the
expected number of edges in a set of vertices of the same degrees as
$K_n(i)$ in a random graph of the same degree sequence as the original
graph $G$. Calculating of $\mathbb{E} M$ is known to be a hard
counting problem~\cite{BC78}, so we have to rely on the same random
sampling as for the $\ccp$-calculation. To save time one can calculate
$\mathbb{E} M(K)$ as the average number of edges within the original
subgraph $K$ at the same time as the $\mathcal{G}(G)$-sampling of
the $\ccp$ calculation. In Fig.~\ref{fig:rad}(d)-(f) we diagram
$\langle\mu\rangle(n)$ for our three example networks. Since the core
is constructed to be highly inter-connected it is no surprise that
$\langle\mu\rangle$ has a peak for small $n$. For the metabolic
network of Fig.~\ref{fig:rad}(f) this peak is small. This is due to
the exceptionally high degrees $\sim 55$ of the core vertices (including
substrates such as H$_2$O, ATP and ADP)---even in the null-model
networks this set of vertices will, for combinatorial reasons, be
highly interconnected. For intermediate $n$ the
$\langle\mu\rangle$-values are of the order of unity, i.e.\ there is
no overrepresentation of edges between vertices at this distance from
the core. But as $n$ increases, $\langle\mu\rangle$ grows to a sharp
peak before it eventually drops to zero. This seems like a rather
ubiquitous feature (at least it is present in almost all networks of
Table~\ref{tab:ccp}). We interpret this as that the periphery has
both the two functions listed above: To a certain distance from the
core (defined by the peak) vertices have similar function and are for
this reason connected (and since such small set of, probably,
low-degree vertices is unlikely to have many interior edges $\mu$
becomes high); beyond this distance the network consists only of
cycle-free branches. This dichotomy---the network in- and
outside of the peak radius---is yet more distinct than the core vs.\
periphery as defined above. On the other hand the outside is
functionally rather trivial and (in all cases we study) smaller than
the inside (we believe the term ``core'' is more apt for smaller
subgraphs). We note that this peak is not trivially related
to the peak in the size of the $n$-neighborhood which is much broader
and occurs for smaller $n$.

\section{Summary and Discussion}

In many networks the properties of vertices are heterogeneously
distributed, similarly one can find subgraphs with very different
characteristics and function. Perhaps the simplest division of a
network is that into a core and a periphery. The core concept has
been used in various senses in the past; typically it is defined as a
subgraph which is most tightly connected~\cite{be:cp,eb:peri} or a
most central~\cite{chung_lu:pnas}. In this paper we use the rather
strong precepts that a core should be both highly interconnected and
central. To quantify this idea, we define the core as the $k$-core
of highest closeness centrality. Then, to measure the strength of the
tendency to have a central and highly connected core, we define a
core-periphery coefficient as the normalized closeness centrality of
the core minus the same corresponding average value for our null-model
(random networks of the same degree sequence as the original
network). Different types of networks have their characteristic
$\ccp$-values: Geographically embedded networks typically have
a positive core-periphery coefficient. We explain this as an effect of
their communication-time optimization. Social network, on the other
hand, typically have slightly negative $\ccp$ values despite their
positive degree-degree correlations. We show that $\ccp$ for model
networks such as the Erd\H{o}s-R\'{e}nyi, Barab\'{a}si-Albert and
Watts-Strogatz models goes to zero (or at least to a very small value)
as the network size increases but, that
one can construct networks with a positive $\ccp$ in the large system
limit. Once the core of a network is found one can construct a
radial image of the network by plotting quantities averaged over the
$n$-neighborhoods of the core vertices as a function of $n$. One such
quantity we study is $\mu(n,i)$---the relative number of edges within
the $n$-neighborhood of $i$ to the expected number of edges in a
subgraph of the same set of degrees in the
null-model. $\langle\mu\rangle$ shows, almost ubiquitously, a peak at
intermediate $n$. We interpret this peak as an effective radius of
the network. Much remains to be done in terms of characterizing the
cores and peripheries of complex networks. We believe this dichotomy
and the radial imagery we present are very useful tools to understand
the large-scale architecture of such networks.

\begin{acknowledgements}
We thank Mark Newman for valuable discussions, and Martin Rosvall,
Michael Gastner, Jean-Pierre Eckman, Holger Ebel, Mark Newman, Ajou
University, Beom Jun Kim, Sungmin Park, Andreas Gr\"{o}nlund, Jonathan
Goodwin, Christian
Wollter, Michael Lokner and Stefan Praszalowicz for help with data
acquisition. This research uses data from Add Health, a program project
designed by J.\ Richard Udry, Peter S.\ Bearman, and Kathleen Mullan
Harris, and funded by a grant P01-HD31921 from the National Institute
of Child Health and Human Development, with cooperative funding from
17 other agencies. Special acknowledgment is due Ronald R.\ Rindfuss
and Barbara Entwisle for assistance in the original design. Persons
interested in obtaining data files from Add Health should contact Add
Health, Carolina Population Center, 123 W.\ Franklin Street, Chapel
Hill, NC 27516-2524 (addhealth@unc.edu).
\end{acknowledgements}

\end{document}